# Weedy Adaptation in *Setaria* spp.:
## VI. *S. faberi* Seed Hull Shape as Soil Germination Signal Antenna

Jennifer L. Donnelly[1], Dean C. Adams[2], and Jack Dekker[3]

[1] Weed Biology Laboratory, Department of Agronomy, Iowa State University, Ames, IA 50011; Email: poppet@iastate.edu

[2] Department of Ecology, Evolution and Organismal Biology, Iowa State University, Ames, IA 50011; Email: dcadams@iastate.edu

[3] Weed Biology Laboratory, Department of Agronomy, Iowa State University, Ames, IA 50011; Email: jdekker@iastate.edu

**Abstract**

Ecological selection forces for weedy and domesticated traits have influenced the evolution of seed shape in *Setaria* resulting in similarity in seed shape that reflects similarity in ecological function rather than reflecting phylogenetic relatedness. Seeds from two diploid subspecies of *Setaria viridis,* consisting of one weedy subspecies and two races of the domesticated subspecies, and four other polyploidy weedy species of *Setaria*. We quantified seed shape from the silhouettes of the seeds in two separate views. Differences in shape were compared to ecological role (weed vs. crop) and the evolutionary trajectory of shape change by phylogenetic grouping from a single reference species was calculated. Idealized three-dimensional models were created to examine the differences in shape relative to surface area and volume. All populations were significantly different in shape, with crops easily distinguished from weeds, regardless of relatedness between the taxa. Trajectory of shape change varied by view, but separated crops from weeds and phylogenetic groupings. Three-dimensional models gave further evidence of differences in shape reflecting adaptation for environmental exploitation. The selective forces for weedy and domesticated traits have exceeded phylogenetic constraints, resulting in seed shape similarity due to ecological role rather than phylogenetic relatedness. Seed shape and surface-to-volume ratio likely reflect the importance of the water film that accumulates on the seed surface when in contact with soil particles. Seed shape may also be a mechanism of niche separation between taxa.

## Introduction

How local adaptation and phylogenetic constraints shape phenotypic diversity and species patterns has been a long standing question in ecology. We are interested in the evolutionary forces responsible for invasive plant seed shape. Species are identified by seed morphology which tends to be conserved between closely related organisms (Werker 1997) and therefore seed shape should function similarly between close relatives. However, Harper (1965) suggested that it is in seed polymorphisms that "the most sensitive reactions of a species to an alien environment are likely to occur," and that "Seed polymorphisms seem particularly likely to be sensitive indicators of evolutionary change in alien invaders.". Jovaag et al. (2012 a, b, c) have revealed sensitive reactions to environmental conditions in the germination and timing of seedling emergence among and within locally-adapted populations of *Setaria faberi*, a prolific weed which has invaded and spread throughout North America (Slife 1954, Warwick 1990). The investigation of whether Harper's conjecture is correct may yield valuable insight into the roles played by phylogenetic conservation and the forces of local adaptation in determining seed shape in invasive species.

The seed is a crucial stage in the life cycle of many plants, particularly those that possess long-term seed dormancy. In many species, seed shape is adapted for dispersal (Peart 1984, Werker 1997, Bekker et al. 1998). However, the function of seed envelope shape in many grasses (eg. *Setaria*, lemma and palea) is not well-known. If Harper's (1965) supposition is correct, the morphology of the outer seed as the primary interface with the environment should differ between closely related taxa depending on local conditions and ecological niche. Therefore, closely related species with differing life-histories should reflect these ecological differences and species sharing a niche should be more similar to one another than relatedness would predict. This would give evidence that selection is acting to drive the evolution of seed shape. However, if phylogenetic relatedness constrains seed shape to be more similar between closely related taxa, then closely related taxa should be more similar in shape to each other than to the taxa which share an ecological niche.

The weedy and domesticated members of the genus *Setaria* (foxtails) are well-suited to examine the question of whether seed shape is determined primarily by function or constrained by phylogeny because congenerics of this species-group, wild-crop-weed complex (de Wet et al. 1979), or polyploid species cluster (Zohary 1965), coexist in close proximity in many habitats (Dekker 2003). Foxtails are one of the worst weed groups in the world, interfering with agriculture and land management (Dekker 2003, 2004) and, therefore, one of the world's most successful invasive plant groups. Many weedy traits expressed over the course of their life histories contribute to their success. One important trait is the production and dispersal of heterogeneous seed, each with a different dormancy capacity (Dekker et al. 1996, Dekker et al. 2001, Dekker and Hargrove 2002a). This heterogeneous seed dormancy inevitably leads to formation of long-lived seed pools in the soil (Jovaag et al., 2012a, b, c), with seedling emergence timing closely tied to initial dormancy capacity (Jovaag et al. 2012c). The ability of *Setaria* seeds to form long-lived seed pools which germinate throughout the growing season and their ability to coexist in close proximity to congenerics for niche exploitation contribute significantly to the success of this species-group. Selection for traits desirable in crops has had a significant impact on the behavior of foxtail millet (*Setaria viridis*, subspecies *italica*), resulting in near-uniform and complete germination and almost simultaneous seedling emergence. The differences in life history between the weedy and domesticated taxa in *Setaria* are likely to be reflected in seed shape if the outer hull plays an important role in the behaviors that differ between weedy and domesticated taxa.

In this study, we examine seed shape in five species of *Setaria* with varying degrees of relatedness. The weedy species are *S. faberi* (giant foxtail), *S. geniculata* (knotroot foxtail), *S. pumila* (yellow foxtail), and *S. verticillata* (bristly foxtail). We also examine two subspecies of *S. viridis* (the weedy *S. viridis* subsp. *viridis* [green foxtail] and domesticated *S. viridis* subsp. *italica* [foxtail millet]). In foxtail millet, we examine two races (*moharia* [moharia millet] and *maxima* [maxima millet]). Giant and bristly foxtail are specialized tetraploid descendants of green foxtail (Dekker 2003). Yellow and knotroot foxtail are also tetraploids and probably share a relatively recent common ancestor (Wang et al. 1995b), although their relatedness to the rest of the genus is considered controversial (e.g.(Rominger 1962, Doust and Kellogg 2002, Dekker

2003)). Because there are no molecular phylogenies of *Setaria* currently available, we have used the phylogeny proposed by Rominger (1962).

The null hypothesis being examined in this study is that phylogenetic relatedness explains the pattern of differences in seed shape between green foxtail and the other taxa being examined (Harper's conjecture does not hold for this species). Our alternative hypothesis is that phylogenetic relatedness is not sufficient to explain the pattern of variation between the taxa being examined (Harper's suggestion merits more investigation).

## Methods and Materials

**Seed Selection**

Seven populations of *Setaria* seeds were selected to be examined in this study (Table 1). The particular populations examined were selected as typical representatives of midwestern U.S. agriculture from the extensive germplasm collection of one of the authors (Dekker) (e.g. (Wang et al. 1995a, b, Jovaag et al. 2012a, b, c)). The representative nature of these particular populations is reinforced by findings that showed that the majority of the world's genetic variation in green, yellow, bristly, and knotroot foxtail is found within Iowa and the U.S. Midwest (Wang et al. 1995a, b). Giant foxtail is nearly genetically homogenous; allozyme analysis revealed that world populations exhibit a single genotype with only one reported exception (Wang et al. 1995b). The taxa selected were also chosen based on phylogenetic relatedness and wide distribution (Figure 1). Foxtail millet is believed to have been domesticated several times from green foxtail (Prasada Rao et al. 1987, Fukunaga et al. 1997, Dekker 2003). Bristly foxtail is an autotetraploid descendent of green foxtail (Rominger 1962, Wang et al. 1995b) while giant foxtail is an allotetraploid of green foxtail and another unknown diploid species (Dekker 2003).Yellow and knotroot foxtails are polyploidy taxa more distantly related to green foxtail. Five of the taxa examined, bristly, giant, green, yellow, and knotroot foxtails, are weeds and were collected in the United States. The remaining two populations, maxima millet and moharia millet, are types of foxtail millet. Because foxtail millet is rarely grown in the US, populations were obtained from Germany and the former Soviet Union, respectively. These seeds were provided by the United States Department of Agriculture – Agricultural Research Service (USDA-ARS), North Central Regional Plant Introduction Station, Ames, IA. Thus, our taxa represent a progenitor weed species, derived crops, weedy polyploid descendants, and more distantly related weeds. Together, these taxa provide us with a framework in which we can examine the nature of selection and phylogenetic constraints in relation to seed shape in *Setaria*.

To verify that the seeds used were representative of their species and as a way to document differences between taxa, size was measured from the lemma using three variables, maximum length (mm), maximum width (mm), and area (mm$^2$). Length was measured from the distal tip of the lemma to the base of the placental pore. Width was measured at the widest part of the lemma and was orthogonal to the length measurement. Area was determined in tpsDig (Rohlf 2004) by measuring the area enclosed by the outline of the lemma. Length-to-width ratios were calculated to give an intuitive, single variable measure of difference in shape.

**Table 1**. *Setaria* accessions by taxa (species, subspecies, race, common name), weed or crop classification, collection location, and accession and germplasm library numbers. [1] J. Dekker

germplasm collection, Weed Biology Laboratory, Agronomy Dept., Iowa State University, Ames, IA 50011 USA.  [2] United States Department of Agriculture – Agricultural Research Service (USDA-ARS), North Central Regional Plant Introduction Station.

|  | Species | Common Name | Weed or Crop | Location | Accession Number |
|---|---|---|---|---|---|
| Diploid | *Setaria viridis* | | | | |
| Diploid | subsp. *viridis* | green foxtail | weed | Ames, IA USA | 3772[1] |
| Diploid | subsp. *italica* | | | | |
| Diploid | race *maxima* | maxima millet | crop | Former USSR | 3764[1]; PI315088, lot 94ncai01[2] |
| Diploid | race *moharia* | moharia millet | crop | Saxony, Germany | 3763[1]; Ames22563, lot 99cai01[2] |
| Polyploid | *Setaria faberii* | giant foxtail | weed | Ames, IA USA | 1816[1] |
| Polyploid | *Setaria geniculata* | knotroot foxtail | weed | Clarkdale, AR USA | 1751[1] |
| Polyploid | *Setaria pumila* | yellow foxtail | weed | Ames, IA USA | 3785[1] |
| Polyploid | *Setaria verticillata* | bristly foxtail | weed | North Platte, NE USA | 1753[1] |

**Shape Collection**

The shape of the silhouette each seed was quantified using geometric morphometric methods (Rohlf and Marcus 1993, Adams et al. 2004). To capture the shape of the silhouette of each seed, two views were digitally photographed: the *dorsal* view and the *lateral* view. For the dorsal view, the seeds were placed on their paleas so that a horizontal plane through the center of the seed was parallel to the work surface. Magnification levels of 40-60x were used to maximize the proportion of the image taken up by the seeds, thus minimizing the variance in measurement due to size. For the lateral view, the seeds were turned to their sides on a piece of fleece fabric so that a plane through the center of the lemma and palea was parallel to the work surface.

Images were collected using a Nikon DXM-1200 digital camera and ACT-1 software (Nikon Inc. 2000).

The outline of the silhouette of the lemma was digitized into a series of landmarks and semilandmarks using tpsDig (Rohlf 2004). Three landmarks and eighteen semi-landmarks were used to capture the shape of the dorsal view and three landmarks and eleven semi-landmarks were used for the lateral view. These landmarks were aligned and converted to size-free shape variables through generalized Procrustes analysis (GPA) in tpsRelw (Rohlf 2003).

**Figure 1.** Relationships of the taxa examined in this study as proposed by Rominger (1962). Scale in the seed images is in millimeters. Note: the branch lengths are not to scale. The exact branch lengths are not known. [Photo: Donnelly, Adams, Dekker]

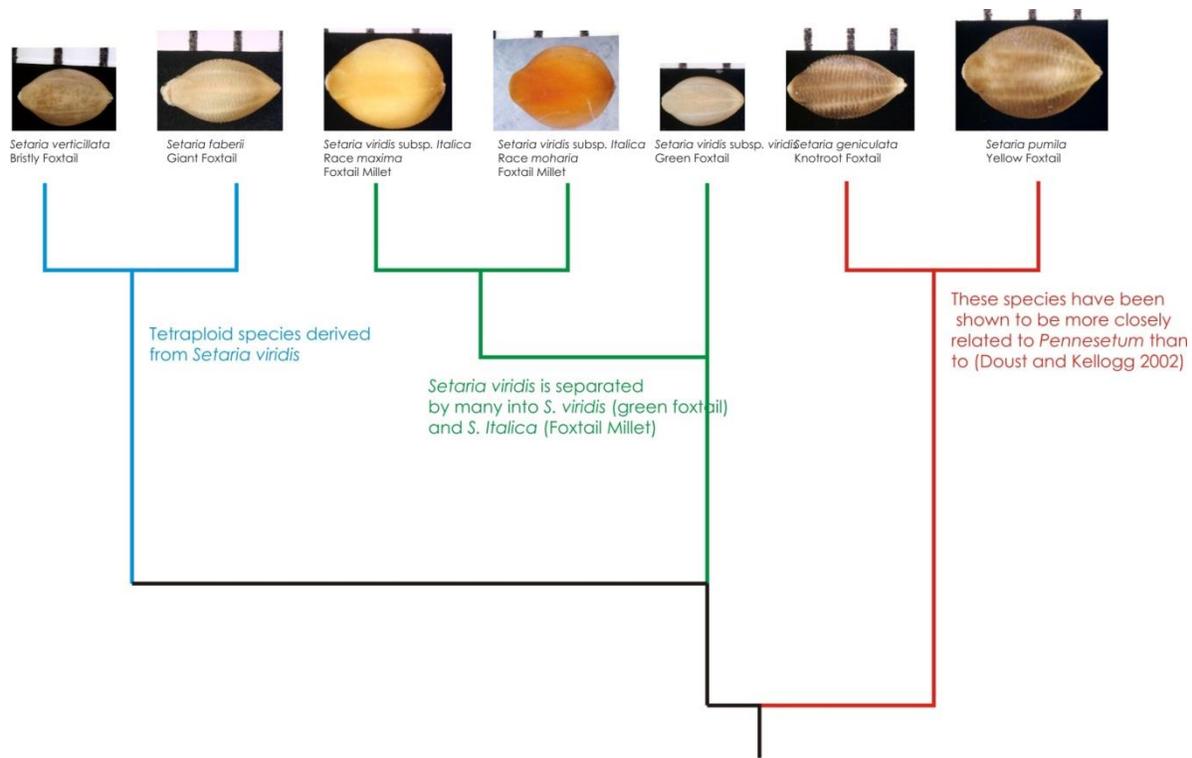

**Analysis**

Differences between length, width, area, and length-to-width ratio were examined by MANOVA performed in JMP (SAS 2002). Pairwise ANOVAs were then performed to determine in which measurements the groups significantly differed.

Dorsal and lateral views of the seeds were analyzed separately. Primary examination of the data was conducted by principal components analysis (PCA) for each view. Canonical variates analyses (CVA) were then performed by view to determine whether there were significant differences between taxa. These analyses (both PCA and CVA) were performed using NTSYSpc (Rohlf 2000). Randomization tests were used to determine whether Mahalanobis distances between groups were significantly different. Each seed was randomly assigned to a group 1000 times and the distances between groups calculated. The percentage of distances greater than those found in this study gave the significance level.

**Analysis of Phenotypic Change**

In addition to more standard analyses of static patterns of phenotypic variation, we also investigated differences in shape between taxa in terms of patterns of phenotypic change, where the phenotypic attributes of various taxa were examined relative to a reference population. Viewed in this manner, one can quantitatively determine whether evolutionary changes in the phenotype were generated more as a result of domestication or phylogenetic divergence. Based on the suggestions of Benabdelmouna et al. (Benabdelmouna et al. 2001), *Setaria* was divided into three groups: the diploids (green foxtail and both races of foxtail millet), tetraploids descended from *S. viridis* (bristly and giant foxtails), and the remaining taxa. This grouping was also consistent with the proposed phylogeny of Rominger (1962). Based on these groupings and phylogeny, we chose green foxtail as the reference taxon, and compared phenotypic patterns among the remaining taxa relative to it. The remaining taxa were then grouped into several evolutionary units: 1) domesticated taxa (= crops, foxtail millet, both races), 2) tetraploid direct descendants (=BG, bristly and giant foxtails), and 3) the more distantly related taxa (= YK, yellow and knotroot foxtails).

**Figure 1**. An example of a *Setaria faberi* seed used in this study in the dorsal view (left) and the lateral view (right). White dots represent the placement of landmarks, (21 for the dorsal view and 14 for the lateral). Maximum length, maximum width, and area were measured from the dorsal view. [Photo: Donnelly-Adams-Dekker]

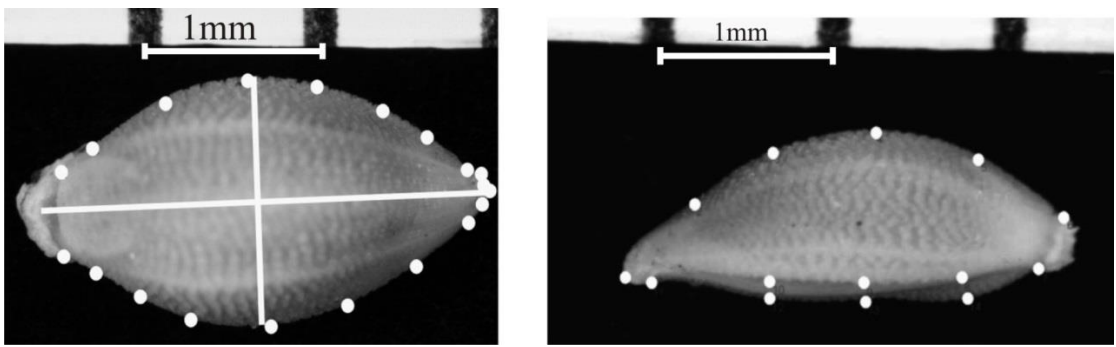

Patterns of phenotypic change among these groups were quantified and compared using the vector comparison approach of Collyer and Adams (2006). With this approach, the average shape for each group, including the reference, was first determined. Phenotypic change for each group was then calculated as the difference in multivariate mean vectors: $\Delta \overline{\mathbf{Y}}_i = \overline{\mathbf{Y}}_i - \overline{\mathbf{Y}}_{ref}$, for each population. From this vector, the magnitude of phenotypic change was calculated as: $D_E = \| \Delta \overline{\mathbf{Y}}_i \| = (\Delta \overline{\mathbf{Y}}_i^T \Delta \overline{\mathbf{Y}}_i)^{1/2}$, where $^T$ represents a vector transpose. Differences in the magnitude of phenotypic change between groups was assessed with the test statistic $| D_{E_i} - D_{E_j} |$, which was calculated for each pair of taxa.. Differences in the direction of phenotypic change between taxa was then calculated as the arccosine of the correlation between normalized vectors of phenotypic change, $\theta$ (Collyer and Adams 2006). The significance of these two test statistics were then evaluated through a randomization procedure where, for each iteration, seeds were randomly assigned to a group, vectors of phenotypic change were recalculated, and the two test statistics were quantified for each pair of taxa (for details (Collyer and Adams 2006).This process was repeated 500 times (including the original sample) to generate distributions of random test statistics to determine whether the observed values were more extreme than was expected from chance.

**Independent Contrasts Analysis**

To verify that shape variation cannot be attributed to phylogenetic relatedness, the data were analyzed using Felsenstein's independent contrasts method (Felsenstein 1985). Molecular sequences used by Kellogg et al (2009) were acquired from TreeBASE (http://www.treebase.org, accession S2357). Because there were only single sequences of *S. faberi, S. viridis, S. pumila, and* foxtail millet (listed as *S. italica* in the data file), those sequences were selected for this analysis. The sequence of *S. verticillata* which was found to be closely related to *S. viridis* by Kellogg et al (2009) was selected because the sample of *S. verticillata* we used was found to be very similar genetically to *S. viridis* (cite). A single instance of *Cenchrus echinatus* (accession AF499151) was selected arbitrarily as an out group. Trees were calculated using Phylip dnaml (Felsenstein 2005) with each of the five accessions of *S. parviflora*, singly and with all accessions included, to determine if choice of sequence affected tree topology. Because the shape data was taken from populations other than those which provided the molecular data, an independent contrasts analysis was performed on the shape data individually for each accession of *S. parviflora* using Phylip contrast (Felsenstein 2005).

**Analysis of Seed Surface Area and Volume**

Recent studies of *Setaria* (Dekker et al. 1996, Dekker et al. 2001) have indicated that seed shape in *Setaria* may be related to the transduction of an environmental signal, which should be reflected in differences in surface area-to-volume ratio between weeds and crops. We created two idealized three-dimensional models of the seeds to examine the relationship between shape and surface area-to-volume ratio (S:V).

We began by assuming the seed is ellipsoid in shape (sensu (Dekker and Luschei 2006). Because there is no closed form for the calculation of the surface area of an ellipsoid, we used a Legendre approximation (Legendre 1825, Tee 2004) with two terms. The standard volume formula for an ellipsoid, $V = 4/3(\pi abc)$ (Weisstein 2005), was used to calculate the volume. To generate the ellipsoids, we used the measured lengths (*a*), widths (*b*), and the maximum height of the lateral view (*c*) as the axes of the shape.

For the second idealized model, the *reconstructed* model, we converted the landmark outlines to continuous curves using elliptical fourier analysis (EFA) in Morpheus (Slice 1998). The lateral view gave the seed shape in the XY plane and the dorsal view gave the shape in the XZ plane. The two views were then scaled so that the reconstructed seed was the length measured in the analysis of size. We assumed the silhouettes of the dorsal and lateral views crossed at the center of the dorsal view and the longest point of the lateral view, giving the lemma above the point of intersection and the palea below. We then assumed that the lemma and palea are each half-ovals in cross section. The major axis of the upper oval was determined by the height of the lateral outline from the point where the two views intersected. Similarly, the major axis of the lower oval was determined by the distance from the intersection line and the outline below it. The minor axes of both the upper and lower ovals were determined by the width of the dorsal view at that point. The half-ovals were drawn with a constant number of points at equally spaced angles per seed for ease of drawing a 3-dimensional mesh. The number of points per oval was chosen by determining the maximum height of the lateral view, then selecting the angle size which would give a maximum distance between points of 0.005 mm. This value was chosen so that error introduced by using points this closely spaced together is insignificant compared to the error introduced by the assumptions of a smooth surface and the shape of the cross-sections of the seeds. The points were converted to a standard stereo lithography (STL) format, which describes the surface in triangles. The surface area is calculated by summing the areas of these triangles. Surface area and volume measurements were collected using ModelPress (Informative Graphics Corp 2004, Informative Graphics Corp. 2004) and the ratio of surface area to volume (S:V) was calculated.

**Figure 2**. The three-dimensional idealized seed was created using two orthogonal views of the seed. The dorsal (XY plant) and lateral (XZ plant) views were assumed to intersect at the point where the seed was the longest. The seeds were assumed to be half-ovals above and below the plane of the dorsal view. The minor axis of each half oval was determined by the width of the lemma at that point. The major axis was determined by the distance from the plane of intersection to the outline of the lateral view. The points were taken so that, when converted to STL format, the triangles formed would have sides no greater than 0.005mm in length.

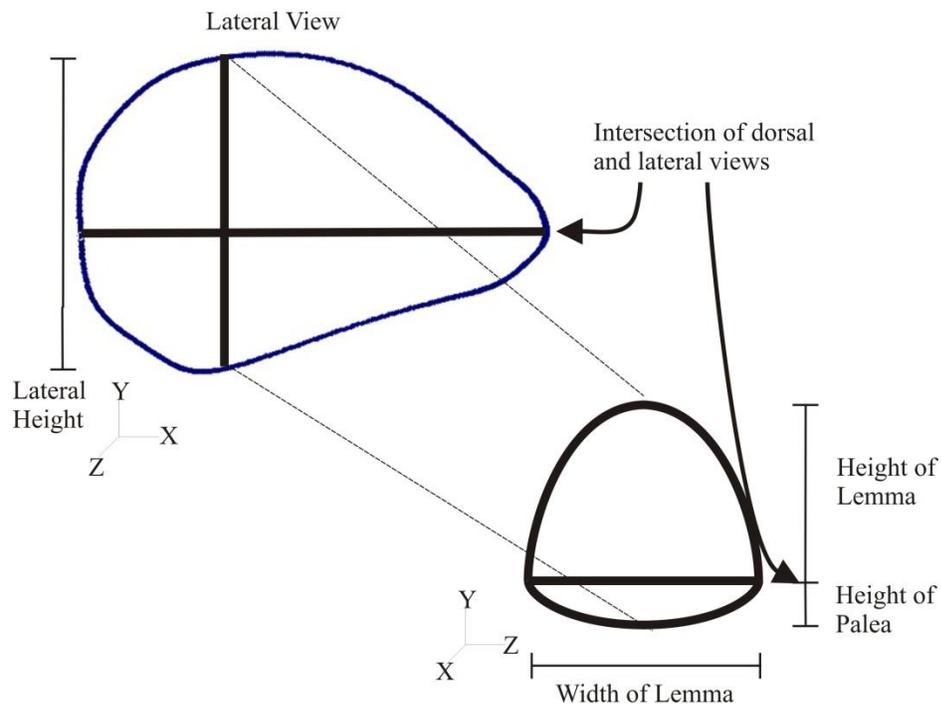

S:Vs were compared using ANOVA and then tested by pairwise ANOVA in JMP (SAS 2002) to determine whether there were significant differences between taxa. Differences between the ellipsoid and reconstructed models were quantified by comparing S:Vs from each model by t-test in JMP (SAS 2002).

## Results

**Size and Appearance**

We began with the assumption that the seed lots selected were representative of the species under examination. To verify this, we compared seed size measurements (length) with those reported in Rominger's (1962) definitive work on *Setaria*. All weed seed lengths fell within the size ranges given for these species except knotroot foxtail, which was the most variable in length. However, the average length of knotroot foxtail found in this study was within one standard deviation of the longest length given by Rominger (2.8mm) so this population was not deemed to be atypical.

Once our assumption of representative populations had been verified, we examined whether there were size and shape differences between the taxa. A MANOVA analysis of size found significant differences between populations ($p < 0.05$). Pairwise ANOVAs (with Bonferroni adjustment for multiple comparisons, $p<0.0006$) were used to determine which populations differed and in which dimension. Giant foxtail and maxima millet did not differ in length, but all other measurements proved to be significantly different between taxa. The weedy and domesticated taxa were also easily separated by average length-to-width ratios, which were significantly different ($p < 0.001$). Weedy taxa had length-to-width ratios between 1.45 and 1.86 while the values for crops were 1.22 and 1.36.

In addition to the differences in size, there were differences in the appearance between weeds and crops. The lemma and palea of all of the weedy taxa were indurate (hard) and the suture between the lemma and palea was tightly fused at all points except at the abscission point which has a transfer aleurone cell layer (TACL) to filter materials entering and exiting the seed (Rost, 1970 #842). In foxtail millet (both races), the lemma and palea were thinner and fragile, cracking easily. The suture between the lemma and palea was not fused at the distal end of the seed, allowing materials to pass into and out of the seed without having to transverse the TACL.

| | Taxon | Length (mm) | Width (mm) | Area (mm$^2$) | Length/Width |
|---|---|---|---|---|---|
| **Diploid** | Green Foxtail | 1.8324±0.0834 (1.8-2.2mm) | 1.1505±0.0527 | 1.5489±0.1229 | 1.5941±0.0646 |
| | Moharia Millet | 2.1784±0.0735 (3mm) | 1.7948±0.0797 | 2.8671±0.1746 | 1.2160±0.0670 |
| | Maxima Millet | 2.5349±0.1023* (3mm) | 1.8682±0.0890 | 3.4955±0.1821 | 1.3610±0.1015 |
| **Polyploid** | Bristly Foxtail | 2.0858±0.0687 (2-2.2mm) | 1.2145±0.0770 | 1.8104±0.1052 | 1.7256±0.1415 |
| | Giant Foxtail | 2.5197±0.0747* (2.5-3mm) | 1.4520±0.0463 | 2.5508±0.1393 | 1.7366±0.0614 |
| | Knotroot Foxtail | 2.8543±0.1155 (2-2.8mm) | 1.5500±0.0689 | 3.0645±0.2431 | 1.8423±0.0403 |
| | Yellow Foxtail | 3.018±0.0834 (3-3.4mm) | 2.0476±0.0642 | 4.3351±0.4459 | 1.4748±0.0419 |

While all seeds exhibited some degree of rugosity, it was clear that the millet seeds are much less rugose than any of the weed seeds. Knotroot and yellow foxtails were the most rugose, followed by giant foxtail. Bristly and green foxtails were the least rugose of the weeds. Moharia and maxima millets had shiny, nearly smooth surfaces.

**Shape**

Shape was quantified in two separate views, the dorsal and lemma views. The dorsal view was captured using three landmarks and eighteen semi-landmarks. Three landmarks and eleven semi-landmarks were used to characterize the shape of the lateral view. A generalized Procrustes analysis of these landmarks produced 28 variables describing the shape of the dorsal view and 24 variables describing the shape of the lateral view for each seed. We found the best two-dimensional representations of the data for each view and through principal components analysis. The first two principal components (PC1 and PC2) together captured 89% and 81% of the variation in the data for the dorsal and lateral views, respectively. In both views, the domesticated taxa appeared to be separated from the weedy taxa. In the dorsal view, this separation was most apparent in PC1. A combination of PC1 and PC2 are required to see the separation in the lateral view. Each taxon appeared to form a cluster within the PCA graphs, but overlap made it impossible to determine whether the differences were significant.

To quantify the differences between groups, a canonical variates analysis (CVA) was performed. The MANOVA portion of the CVA showed that the populations were significantly different ($p \ll 0.001$ for each view). Reclassification revealed that 98% of the individuals (412 of 420) were correctly assigned to their original group for each view: one giant foxtail seed was listed as bristly foxtail, three moharia millet seeds were categorized as maxima millet, and four maxima millet seeds were labeled as moharia millet. None of the misclassifications were made between weeds and crops. Further evidence of differences between taxa is provided by the fact that Mahalanobis distances between the populations were significantly different ($p < 0.01$). When the taxa were divided into three groups (crops; bristly, giant, and green foxtails; and yellow and knotroot foxtails), the distances between the groups were significantly different ($p < 0.001$). Dividing the taxa into crops and weeds, the average distance within each group was significantly smaller than the average distance between groups ($p < 0.01$).

**Figure 3**. A principle components analysis performed on the shape variables from the dorsal view of seven populations of Setaria with shape deformation grids showing the average shape for each population. The first two principal components (A) account for 66% of the variation in the data. The line across part A marks the boundary between weedy and domesticated specimens. Each population formed a tight cluster within the graph, as shown in B. Deformations in the grids in B show where the shape of each population varied from the overall average shape of all seeds examined. It is clear when looking at the shape of *S. viridis* subsp. *viridis* that it is more similar to the other four weeds, shown on the left, than to the two domesticated conspecific taxa shown below it.

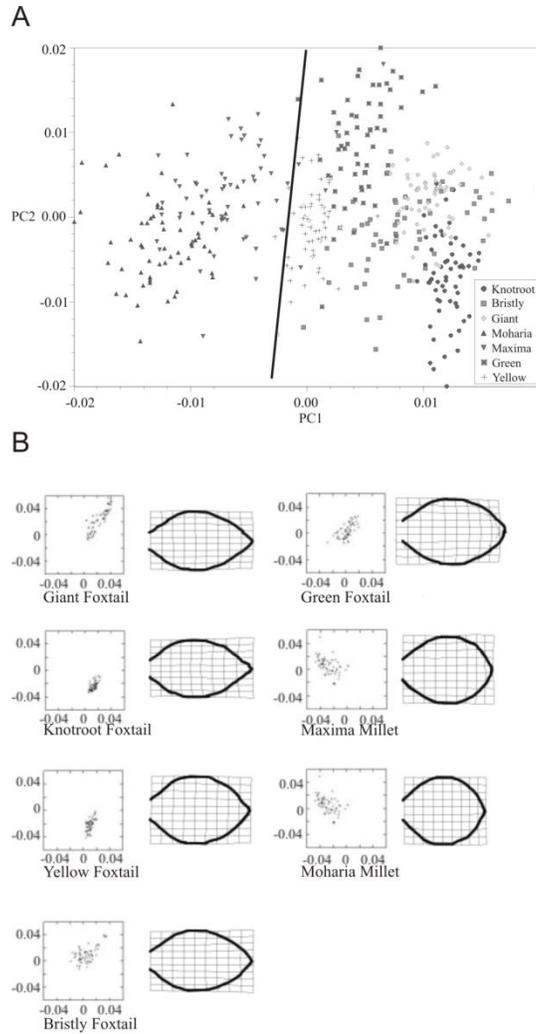

**Figure 4**. A principle components analysis performed on the shape variables from the lateral view of seven populations of *Setaria* with shape deformation grids showing the average shape for each population. The first two principal components (A) account for 66% of the variation in the data. The line across part A marks the boundary between weedy and domesticated specimens. Each population formed a tight cluster within the graph, as shown in B. Deformations in the grids in B show where the shape of each population varied from the overall average shape of all seeds examined. It is clear when looking at the shape of *S. viridis* subsp. *viridis* that it is more similar to the other four weeds, shown on the left, than to the two domesticated conspecific taxa shown below it.

A

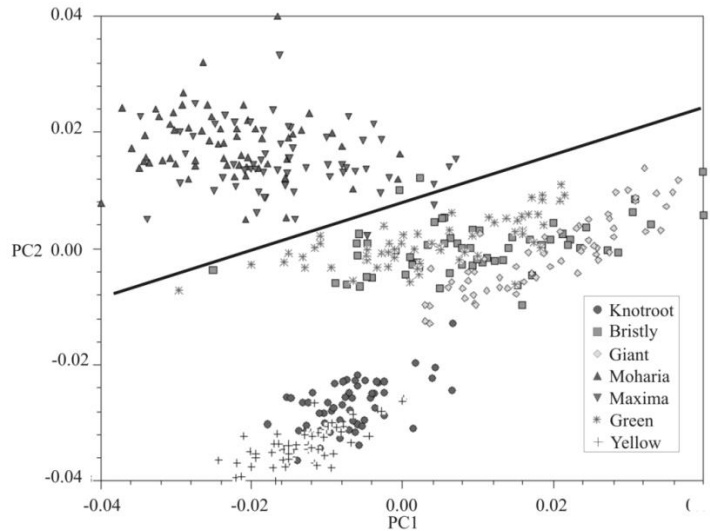

B

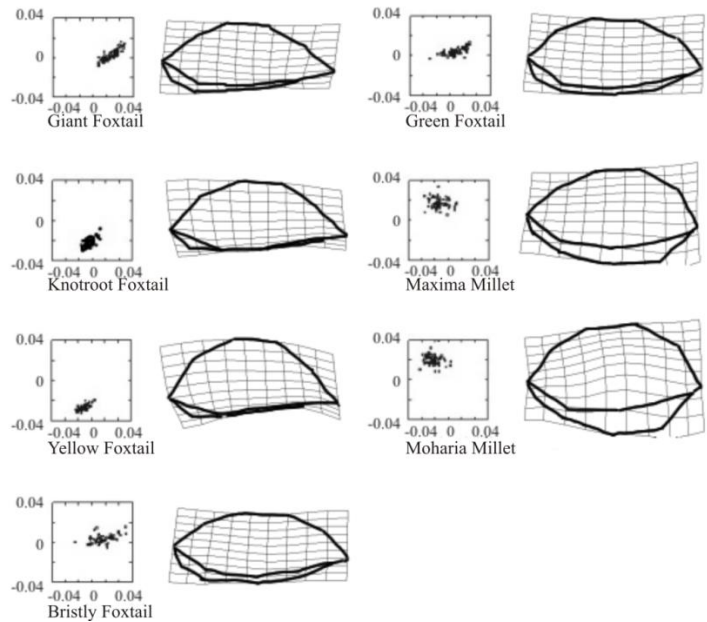

**Analysis of Phenotypic Change**

The dorsal and lateral views gave differing pictures the trajectories of shape change from the reference taxon (green foxtail) for each group.

In the dorsal view, the magnitude of the change vector for the crops significantly differed from both weedy groups (Table 1). The direction of change was significantly different only between the crops and BG. Examination of warp grid representations of the average specimen for each taxon revealed that differences in shape appeared to relate to elongation of the seed). In the lateral view, all trajectories had significantly different magnitudes (Table 1).

**Figure 5**. A principal components plot showing the results of the trajectory analysis for the dorsal view of the seed. Weeds and crops show significantly different magnitudes and directions of shape change from one another. The differences in magnitude and direction of the change for the weedy taxa are non-significant. The reference is *Setaria viridis* subsp. *viridis*. Crops are *Setaria viridis* subsp. *italica* race maxima and *Setaria viridis* subsp. *italica* race moharia. FV is the group of weeds directly descended from *Setaria viridis* subsp. *viridis*, consisting of *S. faberi* and *S. verticillata*. PG refers to the more distantly related foxtails, *S. pumila* and *S. geniculata*.

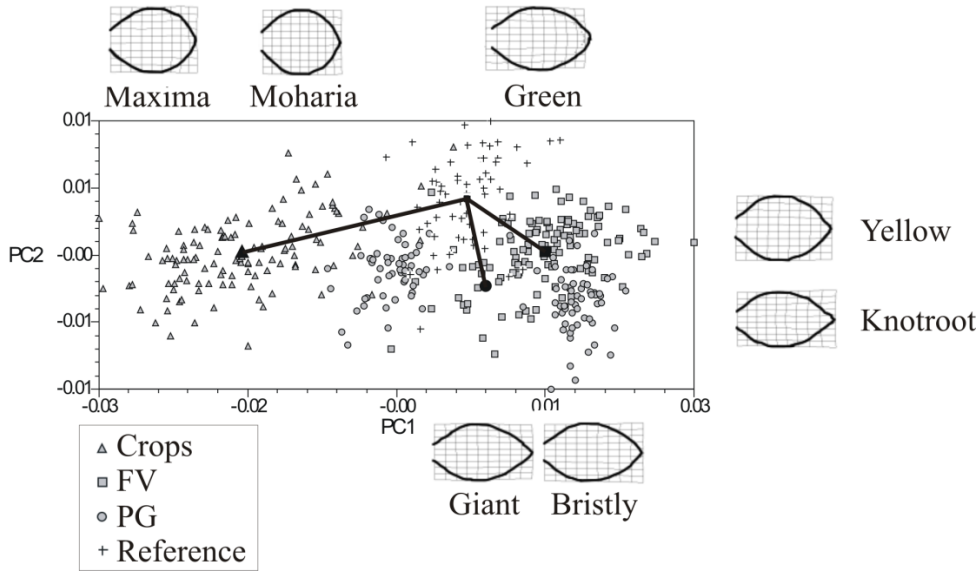

**Table 1**. Significance levels of the comparisons of the vectors of phenotypic change formed by each group from the reference specimen (green foxtail), both in magnitude of change and direction of change. Magnitude is the Euclidean distance of the group mean from the origin after the data were standardized to center at the origin. (green foxtail). Direction is the angle of the vector from the origin to the group mean. The crop group consists of moharia and maxima millets. The BG group consists of bristly and giant foxtails. Yellow and knotroot foxtails make up the PG group. Values in boldface are considered to be statistically significant.

|         | Dorsal View | | Lateral View | |
| --- | --- | --- | --- | --- |
|         | Magnitude | Direction | Magnitude | Direction |
| Crop-BG | **$p = 0.002$** | **$p = 0.026$** | **$p = 0.002$** | **$p = 0.002$** |
| Crop-YK | **$p = 0.002$** | $p = 0.16$ | **$p = 0.002$** | $p = 0.092$ |
| BG-YK   | $p = 0.738$ | $p = 0.422$ | **$p = 0.002$** | **$p = 0.022$** |

**Figure 6**. A principal components plot showing the results of the trajectory analysis for the lateral view of the seed. All magnitudes of the vectors from the reference are significantly

different. FV is significantly different in direction from crops and PG. The difference in direction between the vectors for FV and PG is not significant. The reference is *Setaria viridis* subsp. *viridis*. Crops are *Setaria viridis* subsp. *italica* race maxima and *Setaria viridis* subsp. *italica* race moharia. FV is the group of weeds directly descended from *Setaria viridis* subsp. *viridis*, consisting of *S. faberi* and *S. verticillata*. PG refers to the more distantly related foxtails, *S. pumila* and *S. geniculata*.

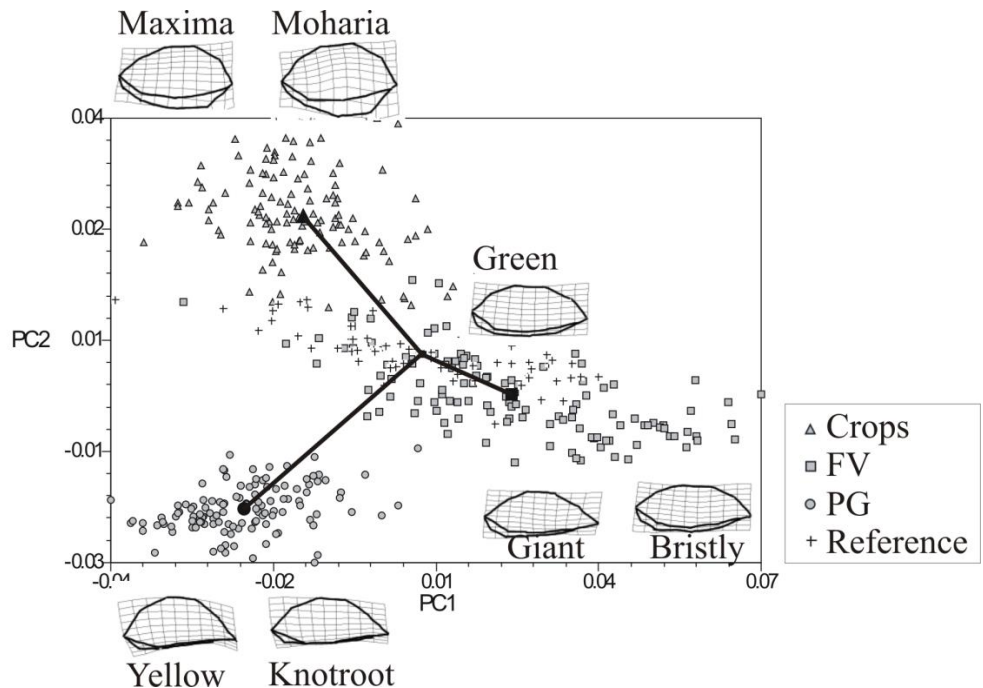

The direction of change was significantly different only between the crops and BG. Examination of warp grid representations of the average specimen for each taxon revealed that differences in shape appeared to relate to elongation of the seed. In the lateral view, all trajectories had significantly different magnitudes (Table 1). The direction of BG significantly differed from the crops and YK, but the direction of change between the crops and YK only approached significance ($p=0.092$). The differences in the lateral view appear to be the result of the differences in shape of the paleas, as seen in the warp grid representation of the average specimen of each taxon. The palea is visible only at the distal end of the seed in YK. In BG and green foxtail, the palea is visible as a thin section along the length of the seed. The paleas of the crops swell below the lemma.

**Independent Contrasts Analysis**

Molecular data gave a phylogeny similar to what was expected a priori (Figure 8). When all five accessions of *S. parviflora* were included, they formed a monophyletic grouping with *S. pumila* (Figure 8a). When examined individually, four of the accessions (AF499142, AY029678,

EU742003, and AF499143)(Figure 8b) were most closely related to *S. pumila*. The fifth sequence (EU7420000) produced a tree very different from the one produced when all the species were included (Figure 8c).

Despite the differences in trees produced, all independent contrasts analyses performed gave similar results. Shape was not correlated with phylogeny ($\chi^2 > 3000$, df=105, $p<<0.005$).

**Analysis of Seed Surface Area and Volume**

The two idealized three-dimensional models gave estimates of surface area, volume, and surface area to volume ratios (S:V) (Table 2) that were significantly different from one another. The values calculated from the reconstructed model were consistently and significantly ($p<0.002$) higher than those calculated from the ellipsoid model. Despite the differences between models, the patterns of differences between taxa were very similar.

Surface areas of all taxa were significantly different in each analysis except giant foxtail and maxima millet in the ellipsoid model. When the taxa were ordered from largest to smallest (Table 2), the two models gave the same ordering for all taxa except knotroot foxtail and moharia millet. Bristly foxtail was larger than moharia millet in the ellipsoid model and smaller in reconstructed. The largest surface areas were calculated for yellow foxtail and the smallest for green foxtail.

**Table 2**. Average seed surface areas, volumes, and surface area to volume ratios (S:V) for two idealized three-dimensional representations of seed shape, ellipsoid and reconstructed from seed outlines. Values from the two models are significantly different according to t-test with Bonferroni adjustment for multiple comparisons ($p < 0.0023$). Means within a column with the same letter are not significantly different as determined by ANOVA with Bonferroni adjustment for multiple comparisons ($p < 0.0023$).

|  | Taxon | Pheno-type | Surface Area | | | | Volume | | | | S:V | | | |
|---|---|---|---|---|---|---|---|---|---|---|---|---|---|---|
|  |  |  | Ellipsoid | | Recon-structed | | Ellipsoid | | Recon-structed | | Ellipsoid | | Recon-structed | |
| Diploid | Green Foxtail | weed | 7.5 | F | 8.2 | G | 0.8 | G | 1.4 | F | 9.6 | A | 6 | A |
| Diploid | Moharia Millet | crop | 13 | D | 14 | E | 1.6 | E | 3.1 | D | 8.4 | B | 4.4 | D |
| Diploid | Maxima Millet | crop | 16 | B | 18 | C | 2.3 | E | 4.5 | C | 7.1 | C | 3.9 | F |
| Polyploid | Bristly Foxtail | weed | 8.9 | E | 9.9 | F | 1.1 | F | 1.8 | E | 8.3 | B | 5.7 | B |
| Polyploid | Giant Foxtail | weed | 13 | D | 14 | D | 1.9 | D | 3.1 | D | 6.8 | D | 4.7 | C |
| Polyploid | Knotroot Foxtail | weed | 16 | C | 18 | B | 2.6 | B | 4.6 | B | 6 | E | 4.1 | E |
| Polyploid | Yellow Foxtail | weed | 21 | A | 24 | A | 3.6 | A | 6.5 | A | 5.9 | F | 3.6 | G |

Similar to surface area, the largest volumes were from yellow foxtail and the smallest from green foxtail. Ordering the taxa from large to small again produced similar results for each model with the exception being giant foxtail and moharia millet. The ellipsoid model gave larger volume estimates for giant foxtail than for moharia millet but the reconstructed model estimates were not significantly different.

The ordering of largest to smallest from S:V were similar to those found from the surface areas and volumes, but inverted. Green foxtail had the highest S:V while yellow foxtail had the lowest. All taxa had significantly different S:Vs for each model ($p \leq 0.0016$).

From examination of the three measures calculated from the three-dimensional models, it was possible to separate the taxa into three groups. The first group, *SMALL*, consisted of green foxtail and bristly. These taxa have the smallest seeds, which gives them the smallest surface area and volumes. However, the S:Vs for these taxa are the highest of all those examined in this study. Phylogenetically, these taxa are very similar, with bristly being the autotetraploid descendent of green foxtail. The second group, *MED*, had intermediate values for surface area, volume, and S:V. Again, these taxa are closely related with the same diploid ancestor, green foxtail. However, giant foxtail also has another diploid ancestor, giving this group more phylogenetic distance than SMALL. The third group, *LARGE*, consists of the taxa that may be the least related to the other taxa, yellow foxtail and knotroot foxtail (Rominger 1962, Doust and Kellogg 2002). These two taxa had the largest seeds with the smallest S:Vs.

## Discussion

In this study, we found evidence that seed shape in the genus *Setaria* is a product of the tradeoff between the constraints of phylogeny and local adaptation. Although there is a general "foxtail shape" (i.e. foxtail seeds from any of the taxa examined are recognizably similar), phylogenetic relatedness alone cannot explain the pattern of shape differences found in this species-group. Green foxtail is more similar in shape to the taxa that share its ecological niche (invasive-colonizing-weedy) than to the crops which are conspecific with it. We found evidence of selection for size, surface to volume ratio, shape, seed surface rugosity, and seed hull permeability to water and gasses (Table 5). From these results, we are able to reject our null hypothesis that phylogeny alone is sufficient to explain the pattern of variation in shape between taxa in *Setaria*. Because there was a clear divide between taxa of differing life histories (weeds vs. crops), we can also reject the possibility that the differences in shape between taxa are random (neutral drift.) Thus, it is likely there is a functional explanation to explain the pattern of variation we have observed.

**Table 5**. Summary of comparison of physical characteristics from taxa examined and groupings as described in the text. S:V and size were calculated in the idealized three-dimensional models (Table 4). Rugosity and water permeability (whether the lemma/palea suture was sealed at all points except at the placental pore) were determined by visual inspection.

| | Taxon | Weed or Crop | S:V | Size (Volume) | Rugosity | Water Permeable |
|---|---|---|---|---|---|---|
| **Diploid** | | | | | | |
| | *S. viridis* | weed | low | small | med | n |
| | *moharia* millet | crop | med | med | very low | y |
| | *maxima* millet | crop | med | med | very low | y |
| **Polyploid** | | | | | | |
| | *S. verticillata* | weed | low | small | med | n |
| | *S. faberi* | weed | med | med | med-high | n |
| | *S. pumila* | weed | high | large | high | n |
| | *S. parviflora* | weed | high | large | high | n |

The literature does not adequately provide an explanation for the pattern of seed shape variation we found. Several roles for seed outer hull surfaces have been proposed, including works on how seed shape is adapted for dispersal (Netolitzky 1926, Boesewinkel and Bouman 1984, Werker 1997), affects orientation of the soil (Peart 1984, Becker et al. 1998), affects persistence in the soil (Thompson and Grime 1979), or is simply an artifact of the space available within the fruit of the parent plant (Kuijt 1967, Lersten and Gunn 1982, Werker 1997) or the shape of the embryo within(Lersten and Gunn 1982, Werker 1997). Not even the comprehensive works on seeds by Baskin and Baskin (Baskin and Baskin 1998) and Werker (1997) provide possible explanations for the pattern of seed shape variation seen in *Setaria*.

Our results can be explained by the hypothesis that seed shape in *Setaria* is adapted to transduce an environmental signal consisting of oxygen, water, temperature, and time (oxy-hydro-thermal time) (Dekker and Hargrove 2002b, Dekker 2003). This signal is used by the seed interior to regulate seed behavior, particularly germination. Moisture availability plays a key role in regulation of germination in giant foxtail (Dekker and Luschei 2009), with too much or too little water inhibiting germination (Dekker and Luschei 2009). With the sensitivity foxtails exhibit to temperature fluctuations, oxygen levels (Dekker and Hargrove 2002a), and volume of water forming a film on the surface of the seed (Dekker and Luschei 2009), we conclude that transducing an oxy-hydro-thermal time signal is a crucial role of the seed exterior. To transduce the signal, the outer surfaces of the seed must accumulate soil water, passively oxygenate the accumulated water, and channel it to the seed interior via the placental pore. In other words, the lemma and palea of a *Setaria* seed act as an antenna to receive, modulate, and transduce information about the environment that the seed interior can use to determine germination.

The efficiency of this antenna is affected by several features of the seed's exterior, size, surface to volume ratio, rugosity, and the permeability of the lemma, palea, and lemma/palea suture to water. The differences we found between taxa show how each group has evolved outer seed morphology to increase the signal reception (Table 5). Green and bristly foxtails have very small seeds, which are more elongate to increase surface area relative to seed volume. Elongation in these species is likely a result of the tradeoff between size and S:V. Giant foxtail, with slightly larger seeds, has also evolved elongation which has resulted in a higher S:V than green and bristly. The seeds of knotroot foxtail are larger and the most elongate with a moderately high S:V. Yellow foxtail has the highest S:V despite being the least elongate and having the largest seed size of the weedy taxa. It may be that life history differences between yellow and knotroot foxtails (knotroot foxtail is the only perennial species in this study) explain the differences in elongation between them. As a perennial, knotroot foxtail has multiple years for reproduction and the potential to reproduce vegetatively (Rominger 1962). This could reduce selection pressure for increased seed volume as found in yellow foxtail. Despite their differences, all the weeds show a distinct selective pressure acting on seed shape. The seeds are elongate to at least a limited extent and all have relatively flat paleas. Foxtail millet's domestication has resulted in much rounder (less elongate), moderately-sized seeds with an intermediate S:V. The palea is extremely rounded in foxtail millet, likely due to selection for maximizing starch content.

Rugosity is likely to amplify the differences seen in S:V in the weedy taxa. Yellow and knotroot foxtails, the taxa with the highest S:V, are also the most visibly rugose. Giant foxtail has more rugose seed surfaces than green or bristly foxtail and also a greater S:V. However, S:V in foxtail millet will likely decrease in relation to the weeds when rugosity is factored in. The seeds of foxtail millet are relatively smooth in comparison to their weedy relatives (Pohl 1951, Rominger 1962, Pohl 1978).

Although it is hard to separate the crops out from the weeds by looking at seed size and S:V, there is a large difference between the taxa in terms of water permeability along the lemma/palea (LP) suture and in seed surface fragility. Because the LP suture is completely fused in the weedy taxa, water must be channeled to the seed interior via the placental pore (abscission point). This is not necessary in foxtail millet because the LP suture is not completely sealed at the distal end and the lemma and palea of foxtail millet are thin and fragile (Pohl 1951, Rominger 1962{Donnelly, #739, Pohl 1978)}.

Another difference between the weeds and crops is germination behavior. The weeds germinate throughout the growing season, adjusting to local agricultural practices (Atchison 2001, Jovaag et al. 2012c). Foxtail millet, on the other hand, has been bred to exhibit nearly simultaneous germination. If seed shape in the weedy foxtails is adapted to transduce the environmental signal the seed interior uses for germination, it is likely the differences in seed shape and water permeability in foxtail millet are due to the release of the selection pressures acting on seed shape. In addition, selective breeding for increased starch volumes in foxtail millet is likely the reason foxtail millets are much more spherical than the weedy taxa (maximizing volume while minimizing surface area).

With the results of this study, the oxy-hydro-thermal time signal regulating seed behavior, sensitivity to water film thickness, and the importance of the outer seed surface as the primary interface between the seed interior and its environment, we are able to reject our null hypothesis that seed shape differences between the taxa in the genus *Setaria* can be explained by phylogenetic relatedness. There is obviously a function to seed shape in these taxa. Differences in seed shape and how seed shape changes with environmental conditions may help explain how these species are able to coexist in the same environment, often growing side-by-side. It is also likely that Harper's (1965) supposition that seed shape is sensitive to environmental differences is correct for this weedy/invasive/colonizing species-group. How seed shape adapts to local environmental conditions within a species as well as between species could give valuable insight into the question of why some grasses are more successful invaders and weeds than others.

## Acknowledgements

The authors would like to thank Dr. Dan Ashlock for creating an atmosphere of interdepartmental cooperation that led to the creation of this study and for his insightful comments. We would also like to thank Dr. James Lathrop for writing the programs to create the reconstructed model from our outlines.

# References

Adams, D. C., F. J. Rohlf, and D. E. Slice. 2004. Geometric Morphometrics: Ten years of progress following the "revolution'. Italian Journal of Zoology 71:5-16.

Atchison, B. 2001. Relationships Between Foxtail (*Setaria spp.*) Primary Dormancy at Abcission and Subsequent Seedling Emergence. M.Sc. Iowa State University, Ames, IA.

Baskin, C. C., and J. M. Baskin. 1998. Seeds: Ecology, Biogeography, and Evolution of Dormancy and Germination. Academic Press, New York.

Becker, R. L., C. C. Sheaffer, D. W. Miller, and D. R. Swanson. 1998. Forage quality and economic implications of systems to manage giant foxtail and oat during alfalfa establishment. Journal of Production Agriculture 11:300-308.

Bekker, R. M., J. P. Bakker, U. Grandin, R. Kalamees, P. Milberg, P. Poschlod, K. Thompson, and J. H. Willems. 1998. Seed size, shape and vertical distribution in the soil: indicators of seed longevity. Functional Ecology 12:834-842.

Benabdelmouna, A., Y. Shi, M. Abirached-Darmency, and H. Darmency. 2001. Genomic in situ hybridization (GISH) discriminates between the A and the B genomes in diploid and tetraploid Setaria species. Genome 44:685-690.

Boesewinkel, F. D., and F. Bouman. 1984. The seed: structure. Pages 567-610 *in* B. M. Johri, editor. Embryology of Angiosperms. Springer, Berlin.

Collyer, M., and D. C. Adams. 2006. Analysis of two-state multivariate phenotypic change in ecological studies. Ecology Submitted.

de Wet, J. M. J., L. L. Oestry-Stidd, and J. I. Cubero. 1979. Origins and evolution of foxtail millets. J. Agric. Trop. Bot. Appl. 26:54-64.

Dekker, J., and M. Hargrove. 2002a. Weedy adaptation in Setaria spp. v. effects of gaseous environment on giant foxtail (Setaria faberi) (poaceae) seed germination. American Journal of Botany 89:410-416.

Dekker, J. H. 2003. The foxtail (*Setaria*) species-group. Weed Science 51:641-656.

Dekker, J. H. 2004. The evolutionary biology of the foxtail (*Setaria* spp.) species-group. Pages 65-113 *in* Inderjit, editor. Weed Biology and Management. Kluwer Academic Publishers, The Netherlands.

Dekker, J. H., B. I. Dekker, H. Hilhorst, and C. Karssen. 1996. Weedy adaptation in *Setaria* spp.: IV. Changes in the germinative capacity of *S. faberii* embryos with development from anthesis to after abscission. American Journal of Botany 83:979-991.

Dekker, J. H., and M. Hargrove. 2002b. Weedy adaptation in *Setaria* spp. V. Effects of gaseous environment on giant foxtail (*Setaria faberii*) (Poaceae) seed germination. American Journal of Botany **89**:410-416.

Dekker, J. H., J. Lathrop, B. Atchison, and D. Todey. 2001. The weedy *Setaria* spp. phenotype: How environment and seeds interact from embryogenesis through germination. Proceedings of the British Crop Protection Conference - Weeds 2001:65-71.

Dekker, J. H., and E. Luschei. 2009. Water partitioning between environment and *Setaria faberi* seed exterior-interior compartments. Agricultural Journal 4(2):66-76. (Chapter 7)

Doust, A. N., and E. A. Kellogg. 2002. Inflorescence diversification in the panicoid "bristle grass" clade (Paniceae, Poaceae): Evidence from molecular phylogenies and developmental morphology. American Journal of Botany 89:1203-1222.

Felsenstein, J. 1985. Phylogenies and the comparative method. American Naturalist **125:** 1-12

Felsenstein, J. 2005. PHYLIP (Phylogeny Inference Package) version 3.6. *Distributed by the author. Department of Genome Sciences, University of Washington, Seattle.*

Fukunaga, K., E. Domon, and M. Kawase. 1997. Ribosomal DNA variation in foxtail millet, *Setaria italica* (L.) P. Beauv., and a sruvey of variation from Europe and Asia. Theoretical and Applied Genetics 95:751-756.

Harper, J. L. 1965. Establishment, aggression and cohabitation in weedy species. Pages 243-268 *in* H. G. a. G. L. S. Baker, editor. The Genetics of Colonizing Species. Academic Press, New York, NY.

Informative Graphics Corp. 2004. ModelPress Desktop, Build 4.3.0.1.1. Informative Graphics, Phoenix, AZ.

Jovaag, K., J. Dekker, and B. Atchison. 2012. *Setaria faberi* seed heteroblasty blueprints seedling recruitment: I. Seed dormancy heterogeneity at abscission. International Journal of Plant Research 2(3): 46-56.

Jovaag, K., J. Dekker, and B. Atchison. 2011. *Setaria faberi* seed heteroblasty blueprints seedling recruitment: II. Seed behavior in the soil. International Journal of Plant Research 1(1):1-10.

Jovaag, K., J. Dekker, and B. Atchison. 2012. *Setaria faberi* seed heteroblasty blueprints seedling recruitment: III. Seedling recruitment behavior. In: Dekker, J., B. Atchison, M. Haar, and K. Jovaag. 2012. Weedy *Setaria* seed life history: Heterogeneous seed rain dormancy predicates seedling recruitment. Lambert Academic Publishing, Saarbrücken, Germany. ISBN: 978-3-8454-7859-3

Kellogg, E. A., Aliscioni, S. S., Morrone, O., Pensiero, J. and Fernando Zuloaga. 2009. A phylogeny of *Setaria* (Poaceae, Panicoideae, Paniceae) and related genera based on the chloroplast gene *ndhF*. International Journal of Plant Sciences, 170(1): 117-131.

Kuijt, J. 1967. On the structure and origin of the seedling of *Psittacanthus schiedeanus* (Loranthaceae). Canadian Journal of Botany 45:1497-1506.

Legendre, A.-M. 1825. Traite des Fonctions Elliptiqes, tome 1. Huzard-Courchier, Paris.

Lersten, N. R., and C. R. Gunn. 1982. Testa characters in tribe Vicieae, with nots about tribes Abreae, Cicereae, and Trifolieae (Fabaceae). US. Department of Agriculture, Technical Bulletin 1667:1-40.

Netolitzky, F. 1926. Anatomie der Angiospermen-Samen, Handbuch der Pflanzenanatomie Bd. X. Borntraeger, Berlin.Nikon Inc. 2000. Act-1.

Peart, M. H. 1984. The effects of seed morphology, orientation and position of grass diasporas on seedling survival. The Journal of Ecology **72**:437-453.

Pohl, R. W. 1951. The genus *Setaria* in Iowa. Iowa State College Journal of Science **25**:501-508.

Pohl, R. W. 1978. How to Know the Grasses. 3rd edition. WCB McGraw Hill, Boston, MA.

Prasada Rao, K. E., J. M. J. de Wet, D. E. Brink, and M. H. Mengesha. 1987. Intraspecific variation and systematics of cultivated *Setaria italica*, foxtail millet (*Poaceae*). Economic Botany 41:108-116.

Rohlf, F. J. 2000. NTSYSpc: Numerical Taxonomy System, version 2.1. Exeter Publishing, Ltd., Setauket, NY.

Rohlf, F. J. 2003. tpsRELW, version 1.35. Dept. of Ecology and Evolution, State University of New York at Stony Brook.

Rohlf, F. J. 2004. tpsDIG, version 2.01. Dept. of Ecology and Evolution, State University of New York at Stony Brook.

Rohlf, F. J., and L. F. Marcus. 1993. A revolution in morphometrics. Trends in Ecology and Evolution 8:129-132.
Rominger, J. M. 1962. Taxonomy of *Setaria* in North America. Illinois Biological Monographs **29**:1-132.
SAS. 2002. JMP: The Statistical Discovery Software. SAS Institute, Inc, Cary, NC.
Slice, D. E. 1998. Morpheus et al.: software for morphometric research. Revision 01-30-98. Department of Ecology and Evolution, State University of New York, Stony Brook, New York.
Slife, F. W. 1954. A new *Setaria* species in Illinois.
Tee, G. J. 2004. Surface area and capacity of ellipsoids in n dimensions. University of Auckland Technical Report 139.
Thompson, K., and J. P. Grime. 1979. Seasonal variation in the seed banks of herbaceous species in ten contrasting habitats. Journal of Ecology 67:893-921.
Wang, R. L., J. Wendell, and J. H. Dekker. 1995a. Weedy adaptation in *Setaria* spp.: I. Isozyme analysis of the genetic diversity and population genetic structure in *S. viridis*. American Journal of Botany 82:308-317.
Wang, R. L., J. Wendell, and J. H. Dekker. 1995b. Weedy adaptation in *Setaria* spp.: II. Genetic diversity and population genetic structure in *S. glauca*, *S. geniculata*, and *S. faberii*.. American Journal of Botany **82**:1031-1039.
Warwick, S. I. 1990. Allozyme and life history variation in five northwardly colonizing North American weed species. Plant Systematics and Evolution 169:41-54.
Weisstein, W. W. 2005. Ellipsoid. MathWorld -- A Wolfram Web Resource.
Werker, E. 1997. Seed Anatomy. Gebrüder Borntraeger, Berlin.
Zohary, D. 1965. Colonizer species in the wheat group. Pages 403-419 *in* H. G. Baker and G. L. Stebbins, editors. Proceedings of the First International Union of Biological Sciences Symposia on General Biology. Academic Press, New York.